\documentclass{aastex}
\usepackage{emulateapj5,times,mathptm}

\def  \CIV         {\ifmmode {\rm C}\,{\sc iv}\,\lambda1549
                     \else C\,{\sc iv}\,$\lambda1549$\fi}
\def  \bOIIIb       {\ifmmode {\rm [O}\,{\sc iii]}\,\lambda5007
                     \else [O\,{\sc iii]}\,$\lambda5007$\fi}
\def  \CVI        {\ifmmode {\rm C}\,{\sc vi}\,367~eeV
                       \else C\,{\sc vi}\,367~eV\fi}
\def  \NV        {\ifmmode {\rm N}\,{\sc v}\,\lambda1240
                       \else N\,{\sc v}\,$\lambda1240$\fi}
\def  \OVI        {\ifmmode {\rm O}\,{\sc vi}\,\lambda0351
                       \else O\,{\sc vi}\,$\lambda1035$\fi}
\def  \MgXI        {\ifmmode {\rm Mg}\,{\sc xi}\,1.34~keV
                       \else Mg\,{\sc xi}\,1.34~keV\fi}
\def  \NVI     {\ifmmode {\rm N}\,{\sc vi}\,0.419~keV
                     \else N\,{\sc vi}\,0.419~keV\fi}
\def  \OVII     {\ifmmode {\rm O}\,{\sc vii}\,0.561~eV
                     \else O\,{\sc vii}\,0.561~keV\fi}
\def  \NeVIII      {\ifmmode {\rm Ne}\,{\sc viii}\,\lambda774
                     \else Ne\,{\sc viii}\,$\lambda774$\fi}
\def  \NeIX      {\ifmmode {\rm Ne}\,{\sc ix}\,\lambda13.7
                     \else Ne\,{\sc ix}\,$\lambda13.7$\fi}
\def  \SiXIII      {\ifmmode {\rm Si}\,{\sc xiii}\,1.85~keV
                     \else Si\,{\sc xiii}\,1.85~keV\fi}
\def  \SiXIV      {\ifmmode {\rm Si}\,{\sc xiv}\,2.0~keV
                     \else Si\,{\sc xiv}\,2.0~keV\fi}
\def  \SiIV         {\ifmmode {\rm Si}\,{\sc iv}\,\lambda1397
                     \else Si\,{\sc iv}\,$\lambda1397$\fi}
\def  \bFeXb       {\ifmmode {\rm [Fe}\,{\sc x]}\,\lambda6734
                       \else [Fe\,{\sc x]}\,$\lambda6734$\fi}
\def  \FeK      {\ifmmode {\rm [Fe}\,{\sc]}\,\lambda1.94
                       \else [Fe\,{\sc]}\,$\lambda1.94$\fi}

\def  \Ka      {\ifmmode {\rm Fe}\,{\sc}\, 6.4~keV
                       \else Fe\,{\sc}\,6.4~keV\fi}

\def\chandra{{\it Chandra}}
\def\asca{{\it ASCA}}

\def\rosat{{\it ROSAT}}
\def\sax{{\it BeppoSAX}}
\def\xte{{\it RXTE}}
\def\xmm{{\it XMM-Newton}}

\def\etal{{\it et al.\/} }

\def\approxlt{\mathrel{\hbox{\rlap{\lower .5ex \hbox {$\sim$}}
        \raise .15 ex \hbox{$<$}}}}
\def\approxgt{\mathrel{\hbox{\rlap{\lower .5ex \hbox {$\sim$}}
        \raise .15 ex \hbox{$>$}}}}
\def\cm{{\rm\thinspace cm}}

\def\Msun{\hbox{$\rm\thinspace M_{\odot}$}}

\def\cmii{\hbox{$\cm^{-2}\,$}}
\def\cmiii{\hbox{$\cm^{-3}\,$}}
\def\cmsq{\hbox{$\cm^2\,$}}

\def\Ux{{$U_x$}}
\def\Uox{{$U_{ox}$}}

\newbox\grsign \setbox\grsign=\hbox{$>$} \newdimen\grdimen \grdimen=\ht\grsign
\newbox\simlessbox \newbox\simgreatbox \newbox\simpropbox
\setbox\simgreatbox=\hbox{\raise.5ex\hbox{$>$}\llap
     {\lower.5ex\hbox{$\sim$}}}\ht1=\grdimen\dp1=0pt
\setbox\simlessbox=\hbox{\raise.5ex\hbox{$<$}\llap
     {\lower.5ex\hbox{$\sim$}}}\ht2=\grdimen\dp2=0pt
\setbox\simpropbox=\hbox{\raise.5ex\hbox{$\propto$}\llap
     {\lower.5ex\hbox{$\sim$}}}\ht2=\grdimen\dp2=0pt

\def\simless{\mathrel{\copy\simlessbox}}

\newbox\grsign \setbox\grsign=\hbox{$>$} \newdimen\grdimen \grdimen=\ht\grsign
\newbox\simlessbox \newbox\simgreatbox \newbox\simpropbox
\setbox\simgreatbox=\hbox{\raise.5ex\hbox{$>$}\llap
     {\lower.5ex\hbox{$\sim$}}}\ht1=\grdimen\dp1=0pt
\setbox\simlessbox=\hbox{\raise.5ex\hbox{$<$}\llap
     {\lower.5ex\hbox{$\sim$}}}\ht2=\grdimen\dp2=0pt
\setbox\simpropbox=\hbox{\raise.5ex\hbox{$\propto$}\llap
     {\lower.5ex\hbox{$\sim$}}}\ht2=\grdimen\dp2=0pt

\def\simless{\mathrel{\copy\simlessbox}}

\def\aa{A\&A}

\def\apj{ApJ}
\def\apjs{ApJ~Supp.}

\def\pasj{PASJ}

\def\etal{et al.}

\def\apj{{\it Ap.~J.}}

\def\mnras{{\it M.~N.~R.~A.~S.}}

\slugcomment{Accepted by The Astrophysical Journal}
\shortauthors{NETZER ET AL.}
\shorttitle{THE X-RAY ABSORBING GAS IN NGC 3516}

\journalinfo{The Astrophysical Journal, ???:???--???, astro-ph/0112027}

\begin{document}

\title
{The Density and Location of the X-ray Absorbing Gas in NGC\, 3516}
\author{
Hagai Netzer,\altaffilmark{1}
Doron Chelouche,\altaffilmark{1}
I.M. George,\altaffilmark{2,3}
T.J. Turner,\altaffilmark{2,3}
D.M. Crenshaw,\altaffilmark{4}
S.B. Kraemer,\altaffilmark{5}
K. Nandra,\altaffilmark{2}
}
\altaffiltext{1}{School of Physics and Astronomy, Raymond and Beverly Sackler
Faculty of Exact Sciences, Tel-Aviv University, Tel-Aviv 69978, Israel.}
\altaffiltext{2}{Laboratory for High Energy Astrophysics, NASA/Goddard Space
Flight Center,  Code 662, Greenbelt, MD 20771.}
\altaffiltext{3}{Joint Center for Astrophysics, Physics Department, University
of Maryland, Baltimore County, 1000 Hilltop Circle, Baltimore, MD 21250.}
\altaffiltext{4}{Department of Physics and Astronomy, Georgia State University,
  Atlanta, GA 30303}
\altaffiltext{5}{Catholic University of America, NASA/GSFC, Code 681,
Greenbelt, MD 20771.}

\begin{abstract}
A new {\it Chandra} observation
and archival observations by \asca\
are used to investigate spectral variations in the Seyfert~1 galaxy NGC~3516
over a period of 7 years.
A large change in flux (factor $\sim 50$ at 1 keV) is observed
between an \asca\ observation in 1994 and the \chandra\  observation
in 2000, with the source  close to the
all-time maximum and minimum X-ray flux states, respectively.
We find the variations in the observed flux and spectra at these 
epochs to be consistent with a constant column density of line-of-sight 
material reacting to changes in the ionizing continuum.
The data from the two epochs are consistent with a simple 
decrease (by a factor 8--10) in the luminosity of a constant 0.5--50 keV slope source 
and a line-of-sight 
absorber with an equivalent hydrogen column density of 10$^{21.9}$ \cmii.
Intermediate
luminosities, sampled during other \asca\ observations, are all fitted by the
same model with a very small change in spectral index 
(well below $\Delta\Gamma = 0.2$).
In addition, analysis of the  long (360 ks) \asca\ observation in 
1998 shows clear
``color'' variations that are entirely consistent with this model and are
interpreted as due to changes in the
opacity of the absorbing gas. The data allow us to put a conservative upper
limit of 60 ks on the recombination time which translates
to  a  lower limit of about $2.4 \times 10^6$
\cmiii\ on the density of the recombining gas and an upper
limit of about $6 \times 10^{17}h_{75}^{-2}$ cm on its distance from the central source. These
are  the best limits  obtained so far on the density and location of the X-ray
absorbing gas in a type-1 Active Galactic Nucleus (AGN). 
They indicate that the absorbing gas
is different, in
terms of its density and location, from  the ionized gas commonly observed
in type-II AGN.  
The \chandra\  ACIS/LETGS data also reveals a strong (EW=290
eV), unresolved 6.4 keV iron line,  a strong \OVII\ line and a
marginally detected \NVI\ line. The former is interpreted as originating in a
large column of gas of lower state of ionization seen in ``reflection'', 
and is consistent with the spectrum at high 
energies at all epochs. 
The two others emission lines are probably emitted by the gas also responsible 
for the line-of-sight absorption.
\end{abstract}


\section{Introduction}
Most low-luminosity Active Galactic Nuclei (AGN) contain significant amounts of
circumnuclear gas which modifies the intrinsic X-ray spectrum by introducing
 strong  absorption features, around 1 keV.
This gas, which  has been named ``Warm Absorber''  or the ``highly ionized
gas'' (hereafter HIG),  must  be photoionized and of moderately low temperature
($\sim 10^5$K), as deduced from many detailed studies. Works by Reynolds, 
(1997), George \etal\ (1998, hereafter G98) and others
show it to be present  in some 50--70\% of all low luminosity type-I
AGN (i.e. broad emission line Seyferts). HIG systems appear to be less common
in higher luminosity type-I objects but the statistics are less secure (George
et al. 2000). Previous attempts to analyze  the gas properties focused on   the
absorbing columns and the level of ionization. The last property is well
described by the X-ray (0.1--10 keV) ionization parameter, \Ux\ (Netzer 1996)
which is the ratio of the 0.1--10 keV photon density to the gas density. 
 G98  showed that while objects containing X-ray absorbing systems vary a lot
in the  column density of ionized gas, there is a surprising uniformity in \Ux\
with most objects  clustering around  \Ux=0.1.

 As yet, there is no real indication of the {\it location} of X-ray 
absorbers.  Plausible arguments have been made both for kpc scale 
and for much smaller distances (e.g. Krolik \& Kriss 1995, 2001; Netzer 2001
 and references therein).
Because of that, there is also no information about the mass of the emitting
material and its  kinetic energy.   The best
constraints  can be obtained by high resolution X-ray spectroscopy yet,
 the fundamental issues are not resolved in the
handful of available sources  (e.g. NGC~5548, see Kaastra et al.
2000; NGC~3783, see Kaspi et al. 2001; MCG-6-30-15, see Branduardi et al. 2000
and Lee \etal\ 2001) despite of the wealth of emission and absorption lines
discovered by the new \chandra\ and \xmm\ observations.

The target of our study, NGC~3516 (z=0.009) is at a distance of 35.7$h_{75}^{-1}$ Mpc
where $h_{75}=H_0/75$ km s$^{-1}$ Mpc$^{-1}$ (Ferruit \etal\ 1998).
This is a moderate luminosity 
(L$_{\rm 2-10~ keV} \simeq 10^{42.5-43.0}h_{75}^{-2}$ erg~s$^{-1}$ during 1994--1998) 
type-I AGN with typical X-ray
properties and a small column 
($N_{H I}^{gal} \simeq 3\times 10^{20}$ \cmii) of Galactic
absorption. The \asca\ based studies of Kriss \etal\ (1996b), Reynolds (1997)
and G98 indicated the HIG has a large  
absorbing column ($N_{H} \sim 10^{22}$ \cmii)
and average
ionization parameter. Detailed observations with \xte\   reveal
large variability on long ($\sim 100$ days) and short (days) time-scales
(Edelson et al. 1999, Markowitz and Edelson 2001).  Nandra et al. (1999)
reported the  presence of a relativistically broadened, variable,
 \Ka\ line.  A recent paper by Guainazzi et al. (2001), based on \sax\
observations,  attributed the X-ray variations during 1996--1997 to large
columns of cold material moving in and out of the line of sight. 

NGC 3516 has been extensively observed in most other wavelength bands.
Most relevant to our study are the UV observations (Walters \etal\ 1990; Kriss 
\etal\ 1996a, and references therein; Crenshaw \etal\ 1999; Huchings \etal\
2001) showing strong, multi-component absorption lines superimposed on the
broad emission lines like \CIV\ and \NV.
 The absorption lines are highly variable  on months to years time scales and
suggest gas  outflow from
the nucleus at a large range of velocities, 0--1300 ${\rm km\ s^{-1}}$. 
Walters \etal\ (1990),   
Kriss \etal\ (1996a) and Mathur, Wilkes and Aldcroft (1997)  
discussed  various aspects of this UV-X-ray relationship. 

This paper reports on new, Cycle 1  \chandra\ observations of NGC~3516. The
source happened to be close to its all-time X-ray minimum and hence the
signal-to-noise (S/N) is not adequate for a detailed study of emission and 
absorption lines.
However, the
very low luminosity provided a unique opportunity to study the time-dependent
properties of the absorbing gas and to obtain the first firm 
limits on its density and location. The new observations are
described in \S2, the physical model and the time-dependent properties are
presented and discussed in \S3\ 
 and a general discussion is given in \S4.

\section{Observations and reduction}

{\it Chandra} observed NGC~3516 on 2000 October 6 for 47~ks with the
Low Energy Transmission Grating Spectrometer (LETGS) in the optical path. 
The ``S-array'' of the 
Advanced CCD Imaging Spectrometer (ACIS) was used as the focal-plane 
detector. A sub-array read-out was employed with a 1.1 sec frame time.
Standard ({\tt CIAO} v2.1)
pipeline spectral extraction  was performed on the data using the corresponding
calibration  files  
({\tt CALDB} v2.6).
The data were screened for known bad pixels and columns and
 ``bad grade'' events were discarded. The screening
 resulted in a net exposure of 44~ks. In order to corroborate our 
results from the 
LETGS $\pm 1$-orders, and to improve our signal-to-noise ratio at 
energies below 1~keV, we have also extracted and analyzed the zeroth-order
spectrum. The  total number of counts is 
6800 for the zeroth-order and 8300 for the combined
$\pm1$ orders.  Background  is negligible for the grating spectrum
and
was not subtracted to avoid corrupting the low S/N low-energy spectrum.
Background has been subtracted from the zeroth-order spectrum. 

Pile-up of  the zeroth-order spectrum, at low
energies,  may become important, despite the
extremely low flux level of the source at the time of observations.
Timing analysis, based on a 500 s binning of the zeroth-order data,
shows a constant count-rate of 0.16 counts per 1.1 sec frame 
throughout the 47 ks observation . Thus uncertainties due to pileup are well
below the other observational uncertainties.
To further test the flux calibration accuracy, 
we have made a very detailed comparison of the
zeroth-order spectrum and the heavily binned 1$^{\rm
st}$-order spectrum.
The two are shown in Fig. 1. The comparison shows a good agreement and
complete consistency within the observational uncertainties 
except for a small discrepancy over the 1--2 keV range.
This discrepancy has already been noted by others (e.g. Turner et al. 2001)
and, despite considerable improvement in instrument calibration, is most
probably the result of residual calibration problems with the  1$^{\rm
st}$-order data. 

\section{The central continuum and the absorbing gas} 

NGC 3516  has been observed in several
optical and IR wavebands (Pogge 1989; Miyaji \etal\ 1992; Ferruit, Wilson and
Mulchaey 1998  and references therein). The central region show extended NLR
emission, on a $\sim 2-10$ arcsec scale, with clear spiral-arms type
structure. X-ray imaging  by \rosat-HRI has
been reported in Morse \etal\ (1995) showing a marginally significant extended
emission in the same general direction as the optical line emission.
Our zeroth-order \chandra\ observations clearly show signs of extended nuclear
emission. This is beyond the scope of the present paper, which focuses on the new
spectroscopic results, and is addressed in a separate publication
(George \etal\ 2002).

\subsection{Spectral analysis of the 1994 and 2000 observations}

Inspection of the new flux-calibrated data (Fig. 2) shows an unusual
X-ray spectrum, with extremely hard continuum for a type-I source and a highly
curved low energy spectrum. There are strong and broad absorption features at
0.6--0.9 keV, two strong unresolved emission lines (\OVII\ and \Ka) and a weak,
noisy emission feature at 0.419 keV identified as \NVI. There is
also a hint of a weak emission line near 905 eV that could be due to
 Ne{\sc~ ix}. While the latter raises the interesting possibility of a highly-ionized
component, the S/N of
this feature is so poor that we neglect it in our analysis.  Table 1
gives a summary of the more important results of the new observations with
their  1$\sigma$ error bars. 
\centerline{}
\centerline{\includegraphics[width=8.9cm]{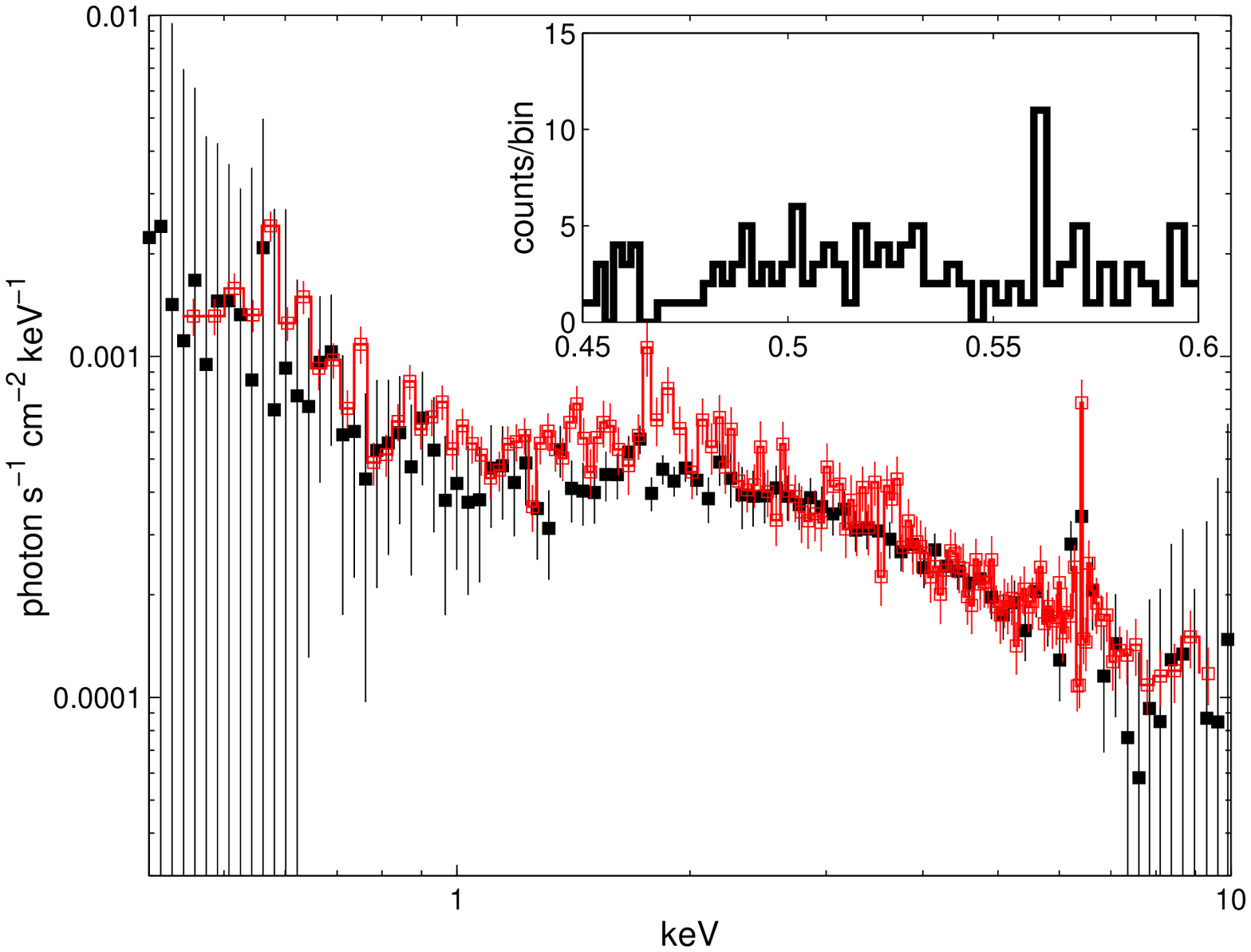}}
\figcaption{A comparison of the zeroth-order (red points) and the
1$^{\rm st}$-order spectra of NGC~3516. 
The combined $\pm 1$ orders data were binned to a constant 
$E/\Delta E  = 30$ resolution. The difference 
between instruments over the 1--2  keV range is  likely due to the known 
LETGS calibration problem. The insert shows the 0.45--0.6 keV range with the
large EW O{\sc ~vii} line at 561 eV. }
\label{LETG-ACIS}

\footnotesize
\begin{center}
\begin{tabular}{ll}
\hline
{\bf Table 1: New \chandra\ results} & \\
\hline
Flux(2--10 keV) & 1.35$\times 10^{-11}$ ergs/\cmsq/sec \\
Flux(0.5--10 keV) & $1.55 \times 10^{-11}$  ergs/\cmsq/sec  \\
EW(\OVII) & $14 \pm 5$ eV \\
EW(\Ka) & $ 290 \pm 50 $ eV \\
F$_E$(continuum at 6.4 keV) & $ (1.6 \pm 0.2) \times 10^{-4}$ photon/\cmsq/sec/keV \\
\hline
Flux(2--10 keV, 1994) & 7.35$\times 10^{-11}$ ergs/\cmsq/sec \\
\hline
\end{tabular}
\end{center}
\normalsize
\centerline
\centerline{}
%
\centerline{\includegraphics[width=8.9cm]{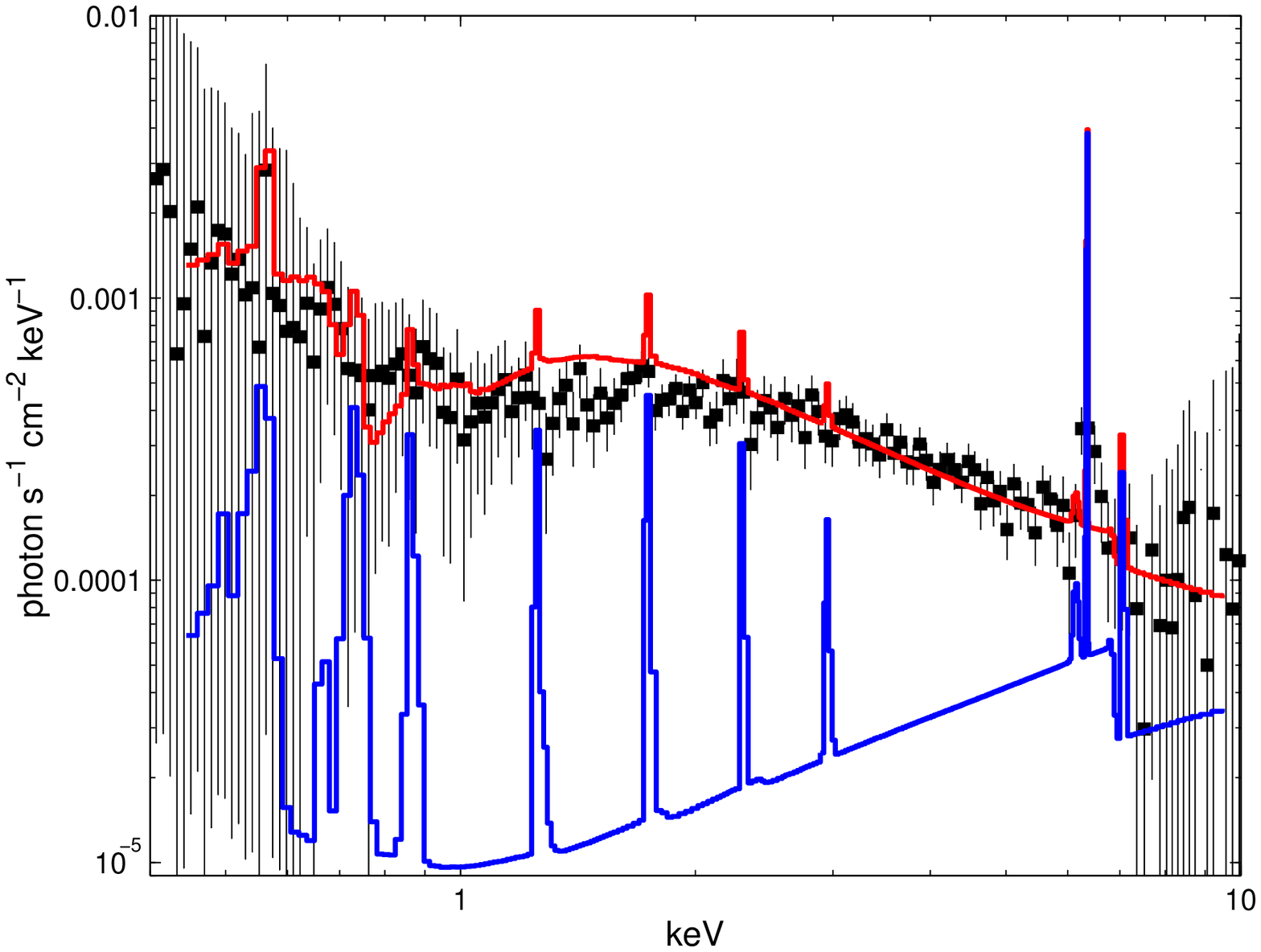}}
\figcaption{ACIS/LETGS spectrum and model fit. The two first-order spectra are
combined to an even resolution of $E/\Delta E =50$ and the model is
rebinned to a similar resolution. The reflection component, with
the assumed normalization of 32\% of the total continuum flux at 6.4 keV, is
shown in blue. The deviation between model and data over  the 1--2
keV range is attributed to the known LETGS calibration problem (see Fig. 1).
}
\centerline{}
\centerline{}

A comparison of the new observations with the 1994-April \asca\ spectrum
suggests a flux drop by a factor  6.4$\pm 0.5$ at 5 keV and a factor $\sim 50$
at 0.7 keV. Thus the spectrum has become harder than almost any other known
type-I AGN spectra. These changes suggest either extremely large intrinsic
slope variations or a large wavelength dependent change in the
opacity of the line-of-sight gas. As shown below, the first of  these
  possibilities seems to
be inconsistent with the available observations.
The following is a
detailed discussion of the second possibility based on detailed modeling of
the central continuum and the absorbing gas.

We have performed detailed photoionization calculations in an attempt to
explain the spectra of NGC~3516 at {\it all epochs}.
Our model calculations use ION2001, the 2001 version of the photoionization
code ION (Netzer 1996; Kaspi et al. 2001 and references therein). The code
includes a self-consistent treatment of the ionization and thermal structure
of a steady-state ionized gas exposed to an external radiation source. Atomic
data include the most recent cross sections for the more abundant elements
and all of their strong lines.
They also include new f-values for a large number of iron L-shell lines 
(see Kaspi et al. 2001), newly calculated 1s-2p transitions (Behar and Netzer 2002)
 and dozens of M-shell iron transitions (Behar, Sako \&
Kahn 2001). These models are more sophisticated than any other photoionization
models applied to this source because of the large number of lines, the inclusion
of both emission and absorption features and the accuracy of the atomic data.

Throughout this work we used models of  constant density ($n_H$)
line-of-sight clouds that are specified  by the continuum
spectral energy distribution (SED), the equivalent hydgrogen 
column density of the gas ($N_H$) and
the ``oxygen ionization parameter'' (\Uox, defined over the 0.538--10 keV
range, see Netzer 2001).  As explained below, fitting of the 4--8 keV
continuum of all X-ray observations (an energy range which is not affected by
absorption)  requires two somewhat different SEDs with 
photon indices  of $\Gamma(0.5-50 ~{\rm keV})=1.7$ and
$\Gamma(0.5-50~{\rm keV})=1.9$. The assumed SEDs are
specified by  broken power-laws between the energies of 0.01, 0.1, 0.5 and 50
eV. In both cases $\Gamma$(0.01--0.1 ~{\rm keV})=2. For the $\Gamma
(0.5-50~{\rm keV})=1.7$  SED case, we have $\Gamma$(0.1-0.5 ~{\rm keV})=3.3
and for the $\Gamma(0.5-50 ~{\rm keV})=1.9$  SED, we need  $\Gamma$(0.1-0.5
~{\rm keV})=3.0. The poor S/N below 0.5 keV does not allow clear determination
of slope (or SED shape) at those energies and the spectral indices over those ranges 
 represent, for example, the  smooth exponential 
distribution of an EUV source. 
 For clarity and references to other commonly
used definitions of the ionization parameter we note that for the 
$\Gamma(0.5-50~{\rm keV})=1.9$ case, $U(13.6 ~{\rm eV})$:\Ux:\Uox=178:13.1:1
and for the $\Gamma(0.5-50~{\rm keV})=1.7$ SED, 
$U(13.6 ~{\rm eV})$:\Ux:\Uox=240:15.4:1,  
where \Ux\ is defined over the 0.1--10
keV range and $U(13.6 ~{\rm eV})$ over the hydrogen Lyman continuum.

The overall  normalizations of the above SEDs  are based on two observed
points, at 1470\AA\ (Kraemer \etal, 2002) taken on 2000 October 1, and near 7
keV (our own observations) taken 5 days later. We assume that this UV-X-ray
flux ratio  has not changed throughout the 7-year period and defer the more
detailed discussion to Kraemer \etal\ (2002). In the X-ray band, the model results are not very
sensitive to this constant-SED assumption since the ionization of the X-ray
absorbing gas is insensitive to the UV radiation. While the chosen 7 keV point 
is not affected by absorption, this is not the case for the
1470\AA\ point which is affected by reddening. This issue has been discussed 
by Kriss \etal\ (1996a) in their analysis of the 1995 observations of the
source. Our adopted value corresponds to E$_{\rm B-V}=0.1$ mag., and is
somewhat larger than the value adopted by Kriss \etal. It is based on the
unusually curved optical-UV continuum of NGC~3516 as observed by HST in 1998
(see Edelson \etal\ 2000) and may, in fact, represent a  lower limit to the
line-of-sight reddening. 
A  more detailed discussion of the SED  and   the issue  of
reddening is  deferred to a future publication (Kraemer \etal\ 2002).

The following analysis is based on fitting the \asca\ and \chandra\ spectra
using standard $\chi^2$ minimization. Throughout this part we adopt the 
acceptability criteria suggested by G98 i.e. a model is considered to be an
adequate description of the data if P($\chi^2 | {\rm dof}) \simless 0.95$. We note that two
of the model parameters,  the covering fraction of the emitting gas and the
FWHM of the absorption lines, cannot be constrained by the data yet they can
affect the calculated $\chi^2$. After testing several cases, we
 have  fixed those to be a covering fraction of 0.4
and pure thermal velocity for the lines.
The  determination of these parameters must
await better S/N observations.
 
We started by  re-analyzing the data obtained during the 
brightest flux state, that obtained from the 
 \asca\ observations in 1994. There is a small flux change during the
observations and our analysis refers only to the first 45~ks 
representing the highest flux.
 The fit result is very similar to the
one obtained by G98 (model B(I) table 5 in their paper). 
It requires a single
full line-of-sight covering HIG cloud with a column density of  $N_H
=10^{21.9}$ \cmii,    \Uox=10$^{-2.1}$ and the  $\Gamma(0.5-50~ {\rm
keV})=1.9$ SED.
We assumed a  neutral
galactic column of 3$\times 10^{20}$ \cmii (see G98) and constant gas density
of  $n_H=10^8$ \cmiii. As argued below, this is consistent with our lower limit
estimate of the HIG density. None of the following conclusions depends much on
the precise value of $n_H$.
Tests show that the uncertainty on the 0.5--50 keV spectral
index is well below 0.2.

The initial fit of the \chandra\  observations was based
 on the assumption of no change in SED and a large luminosity drop 
of the source as inferred from the hard X-ray observations.
 The basic premise is that all spectral variations are due entirely to
flux variations that take place on a time scale much shorter than the seven
year period between observations. The fit obtained in  this way compares
poorly with the observations due to a large excess of flux  around
5--7 keV not represented in the model. This can be explained by the fact
that we also observed a  strong, unresolved  ($ \sigma < 6600$ km/sec)  \Ka\
line with  EW$ =290 \pm 50$ eV. This line intensity is clearly too strong to
be emitted by  the absorbing gas at its 2000 level of ionization and must
therefore represent emission from another, unrelated component. The most likely
explanation is that the line is due to emission by  low ionization gas
(i.e.  much less ionized than the HIG), such as the one commonly
observed by ``reflection'' in type-II (narrow emission lines) AGN. The typical
\Ka\ EW in type-II  sources is about 1 keV. The theoretical value 
(Matt, Brandt \& Fabian 1996; Netzer, Turner \& George 1998) is  also $\sim1$ keV
and depends on the gas composition, the continuum slope and the viewing angle
of the reflector. For the chosen SED, we expect an EW of
700--1000 eV relative to a pure scattering continuum assuming  solar
metalicity and intermediate Compton depth reflector. Given the observed 
\Ka\ line in NGC~3516, we suggest that its origin is the low ionization reflecting
gas and the observed-vs-predicted EW suggests that reflection from this
component contributed between 30 and 50 percent to the continuum at 6.4~keV 
during the \chandra\ observations.
Thus, the factor 5 decrease in the continuum flux {\it observed} at 6.4~keV 
implies a decrease in the {\it true} luminosity of the central source by a factor 8 
to 10 (if 30\% or 50\% of the observed continuum is due to 
reflection, respectively).

To further test the idea of a low ionization component to the \Ka\ line, we 
calculated the emission and reflection spectrum of gas with a large column density  
($N_H=10^{24}$ \cmii) and \Uox=$10^{-4.1}$. Since the
location of this component is probably several light-years from the center,
and the variability time scale of the central source is several weeks or less,
 we had to make an additional assumption about the time-averaged flux level of
NGC~3516. The present model assumes this  to be similar to the high 1994 flux
level. Given this level, the luminosity of the reflection component  requires
a  covering fraction  of   
$\Omega /4 \pi \simeq 0.4 - 0.7 $;
 a reasonable fraction given the likely geometry.
 The emission and scattering  from this gas must be
included when fitting the spectrum at all other epochs. 
In particular,  its contribution is predicted to be non-negligible even at
maximum luminosity. We have therefore included it in our final model of
the 1994 spectrum and, while the overall fit results are unchanged in terms
of SED and column density,  we
note that between 6 and 10 percent of the flux  at around 6.4 keV is due to
this component. The fit results are given in Table 2.

\vspace*{6mm}
\footnotesize
\begin{center}
\begin{tabular}{lcccc}
{\bf Table 2: fit parameters} & & & & \\ 
\hline
Observation &       $\chi^2_{\nu}$ &      d.o.f. &    \Uox\ & $\Gamma$(0.5--50 keV) \\
\hline
ASCA 94 (high)  &       1.07 &    432  &      -2.1  & 1.9  \\
ASCA 95 (high)  &       1.06 &    429  &      -2.3  & 1.9  \\
ASCA 98 (high state)&  0.96  &  373    &     -2.5  & 1.7    \\
ASCA 98 (low state) &  0.68  &  373    &      -2.6  & 1.7  \\
Chandra 0th order   &   0.91 &   263   &      -3.2 &  1.7  \\
Chandra ACIS/LETGS  &   0.48 &   1932  &      -3.2  & 1.7  \\ 
\hline
\end{tabular}
\end{center}
\normalsize
\vspace*{3mm}

Given this additional reflection component, with scattered continuum and
unresolved \Ka\ line, 
we have attempted another spectral fitting of the \chandra\ observations
with the above SED. The fit is statistically acceptable (see Table 2) suggesting that
the reflection component is, indeed, very important for explaining the spectrum
at low flux levels. For reasons related to the 1998 \asca\ observations (see
below) we have also attempted a fit with the  $\Gamma (0.5-50~{\rm keV})=1.7$
SED. This continuum is also consistent with the 2000 \chandra\ observations.
Fig. 2 shows the $\Gamma (0.5-50~{\rm keV})=1.7$ SED model 
on top of the new observations. Inspection of the diagram,  and standard
statistical analysis (Table 2)
 clearly demonstrate the applicability of this approach  
(note however the discrepancy over the 1--2 keV range are due to the known LETGS
calibration feature).
In particular, we conclude that the very
large flux drop, by a factor of $\sim 50$ below 1 keV 
between 1994 and 2000, is entirely consistent with the
constant SED continuum  and the dramatic increase in the
absorption opacity due to the decreasing level of ionization. 

The noticeable flux variations correspond, according to the model, to a
very large change in fractional ionization of the gas.
While the strongest
absorption features in 1994 are due to O{\sc vii} and O{\sc viii}, most of the
opacity in the  0.6--0.9 keV energy range, in 2000, is due to O{\sc v}, O{\sc
vi} and O{\sc vii}.  According to the model, the continuum role-over below 0.6
keV is mostly due to $Z<8$ elements, in particular carbon and helium. In
addition, the model successfully predicts the observed \NVI\ and \OVII\ 
emission line intensities, given the  observational uncertainties and assuming
 the above mentioned covering fraction by HIG clouds with properties similar to those of the
line-of-sight material. This by itself  is not  a conclusive evidence for the
absorber's  properties since the low ionization reflecting component contribution
to the \OVII\ line  can approach the observed line intensity given the
uncertain level of ionization of the reflecting gas.

\subsection{Spectral analysis of the 1995 and 1998 observations}
The seven-year history of \asca, \sax\ and \chandra\ observations of
NGC~3516 provide a unique opportunity to test the suggestion that {\it all}
observed spectral variations are caused by  a single variable,
the unattenuated X-ray flux. To this end, we have collected most of the
previous \asca\ observations of the galaxy and fitted, each one, in the same
way as described   for the  2000 data. The only data set not to be analyzed is from
December 1999. These data suffer from severe calibration problems at low
energies where accurate  spectra are required to test our model.

The results of fitting the \asca\ data  are as follows: 
\begin{description} 
\item[The 1995 data set:] The
mean 0.7--10 keV 1995 luminosity  is about a factor 1.5 lower than the  1994
luminosity and the analysis refers to the first, higher flux  part of the
observations to avoid the continuum variations. We
have kept the $\Gamma$(0.5--50~keV)$=1.9$ SED and scaled \Uox\ by this
factor to obtain a satisfactory fit (Table 2)  to the 
data. Thus the observed 1995 spectrum is  entirely consistent with our
previous assumption.  %
\item[The 1998 data set:]
NGC~3516 was monitored continuously for about 360 ks during 1998 (see
Nandra \etal\ 1999 for details of this campaign). 
The average 6.4 keV continuum flux 
was about 2.5 higher than the \chandra\ 2000 flux, 
suggesting   an intrinsic hard X-ray continuum some  3--4
times  
 more luminous than the 2000 continuum.
 The observations show a large (50\%) peak-to-peak
variations on a time scale of several hours.

Our attempt to fit the mean 1998 spectrum with the 
$\Gamma (0.5-50~{\rm keV})=1.9$ SED 
and the same obscuring column, yielded
unsatisfactory results. We found that, depending on reflector luminosity, the
$E>2$ keV   slope is  flatter than in 1994 and 1995, with $1.7 < \Gamma <
1.8$. The result is unchanged  when splitting the data into high and low
states.   We  conclude
that the source has undergone a real change in slope but by no more than $\Delta \Gamma =0.2$,
between 1995 and 1998 (we re-emphasize that the 1994 0.5--10 keV continuum is
significantly steeper than $\Gamma=1.7$). However, assuming the 
$\Gamma (0.5-50~{\rm keV})=1.7$ SED 
and normalizing \Uox\ by the continuum flux ratio at 6.4 keV between 1998 and
2000,   we can obtain
satisfactory fits to the  high and low 1998 spectra.

The small change of slope at high energies in 1998 is
the main motivation for considering two, slightly different SEDs for
fitting the 2000 observations. The quality of the \chandra\ spectra is not
good enough to distinguish between the two since both give satisfactory
fit to the observations. However, we should note that
the  success of both models   depend  crucially on the assumed 0.1--0.5
luminosity since the (observationally unknown) flux in this band
determines the opacity through the ionization of the $Z<8$ elements. The
assumed 0.1--0.5 keV slope for both SEDs is a direct result of those
constraints and the need to adequately fit the data above 0.5 keV.
\end{description}
 
 To summarize, our model fittings suggest 
$\Gamma$(0.5--50 {\rm keV})=1.9 in 1994 and 1995 becoming flatter, by 
$\Delta \Gamma = 0.1-0.2$
in 1998.  The 2000 observations are consistent with either one of those
SEDs.
Fig. 3 shows our best models and  data-to-model ratios
 for all  data sets (1994, mean-1995, low-1998 and 2000).
Table 2 gives the fit parameters and the calculated
$\chi^2_{\nu}$, all in agreement with our acceptability criteria.   
\begin{figure*}[t]
\hglue0.0cm{\includegraphics[angle=0,width=8.9cm]{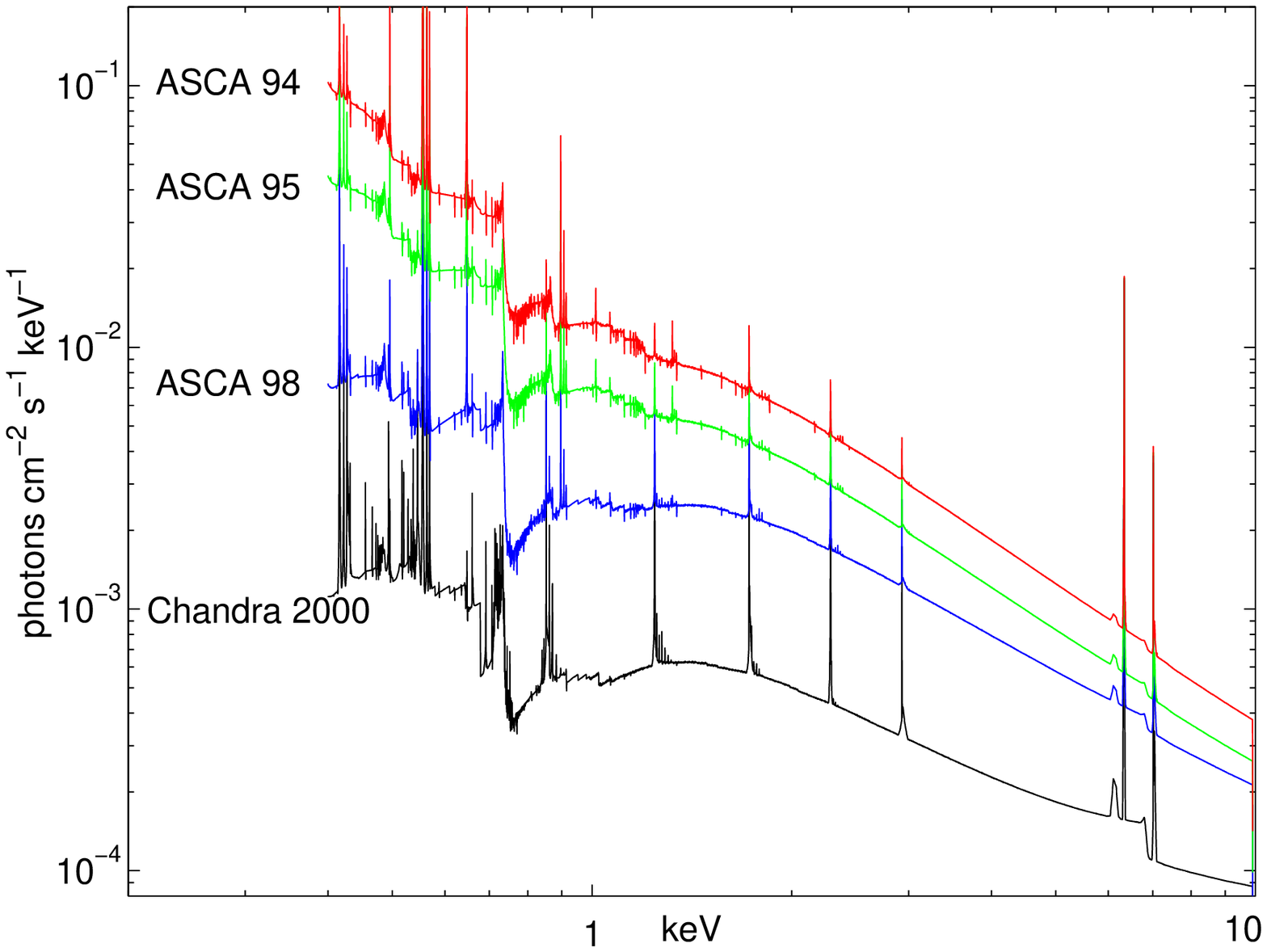}}
\hglue0.5cm{\includegraphics[angle=0,width=8.9cm]{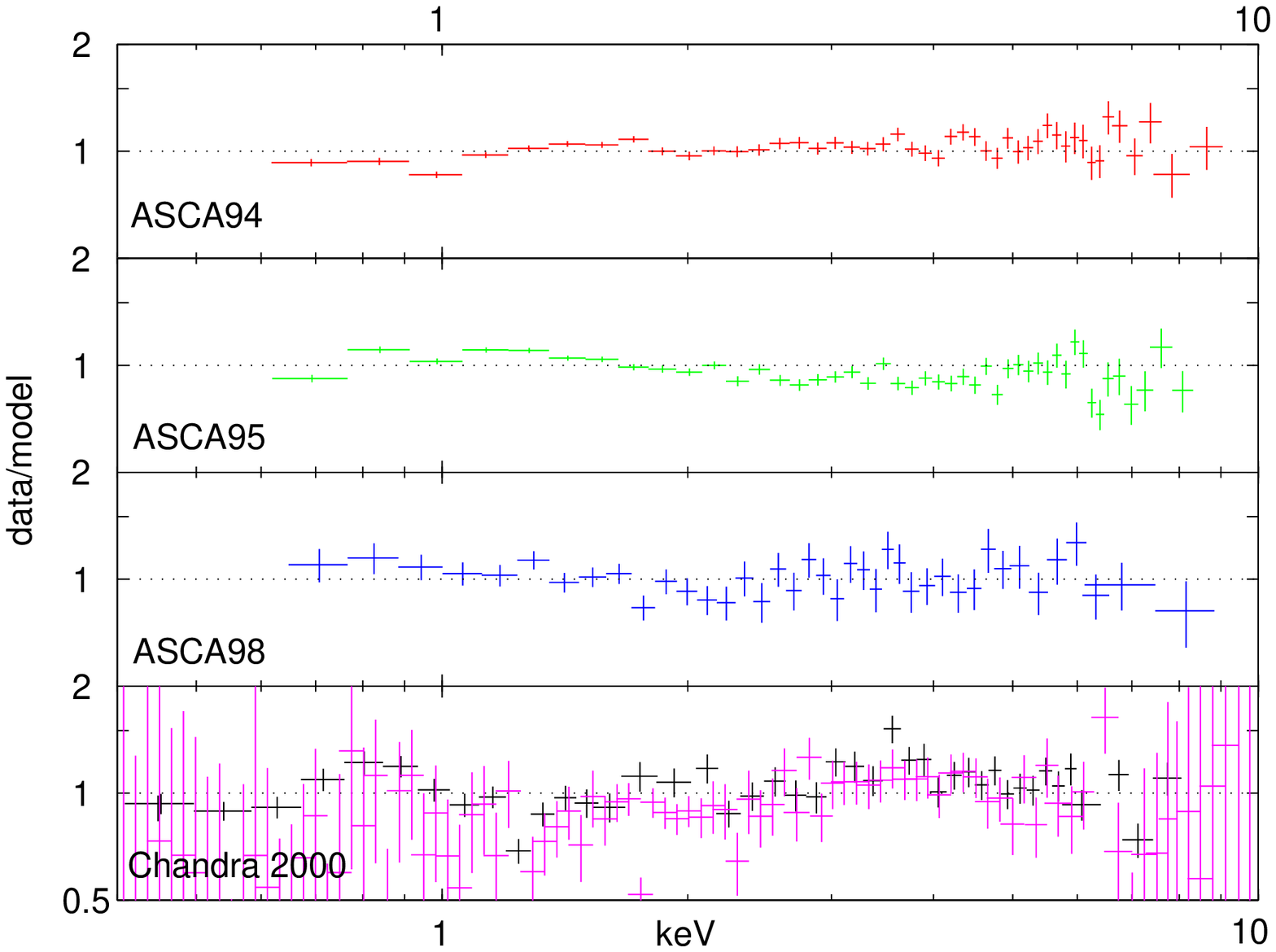}} 
\caption{ {\bf Left:} Adopted models for the 1994 (high level), 1995
  (high level), 1998 (low level at t=300 ks, see Fig. 4) and 2000 observations.
 All models assume $N_H=10^{21.9}$ and the scaling of \Uox\ is according
 to the intrinsic continuum luminosity (see text).
  {\bf Right:} data-over-model ratios for the models shown on the left. Both
 ACIS/LETGS data (in magenta) and the zeroth-order spectrum are shown for the
 \chandra\ observations  
}
\end{figure*} 

\subsection{Color Analysis of the 1998 Data}
Given the satisfactory spectral fits, we can now look for the shortest
variability time corresponding to a detectable change in the gas level of
ionization.  The 1998 data set is most suitable for this since the high energy
continuum varied substantially during the observations.
  Unfortunately, it is difficult to obtain  
meaningful constraints on the time scales  involved using complete spectral fitting 
since it requires integrating over at least 20-30 ks; a period
which is long considering the observed variations in 1998.  Therefore we
adopted the following approach: we have fitted the spectra at two epochs representing
local minima and maxima separated by 60 ks 
(see details below) and we have also used  a complementary
 method based on broad-band photometry. The broad-bands approach 
follows closely the ``color analysis'' method discussed in Netzer,
Turner \& George (1994) and shown to be useful for analyzing absorbed X-ray spectra. 
The method allows the analysis of much shorter data trains with much improved S/N
and considerably better time resolution.
We chose two bands, XM1 which gives the integrated flux over the
0.7--1.2 keV range and XM3 that covers the 4--10 keV range. The fluxes in
these band were measured for the entire 1998 data set in bins corresponding to
one orbit ($\sim 4000$ sec on target integration) and the  ``color''
(or ``softness ratio'') XM1/XM3 was calculated for each of those .  The
results  are presented  
as a function of time,  in Fig. 4.

Inspection of Fig. 4 demonstrates the close agreement of the  XM1 and XM3 flux
variations.  Cross correlation analysis shows the two bands to be highly
correlated, with no measurable time lag, confirming the earlier result of
Edelson \etal\ (2000). This tight correlation  must be
due to intrinsic continuum flux variations.   
The noticeable changes in color, over time scales of a few orbits, are
also very clear.   We suggest that they are caused by  
real changes in the absorber's opacity because of the following reasons:
First,  we have verified, by detailed spectral fitting, that the change of hard
continuum spectral index between  minimum and maximum flux levels during 1998 
is less than $\Delta \Gamma = 0.1$. This  corresponds to a too small change (factor
of less than 1.1) to be responsible to the observed variations in XM1/XM3
(factor about 1.3).
Second, our model calculations (see Fig. 4) are entirely consistent with the
variability amplitude of XM1/XM3 confirming  that
the changing opacity is the main driver for the observed changes.
Cross correlation analysis suggests  a short delay between the
XM3 and the XM1/XM3 light curves but this is not statistically significant.
   
We conclude that a
significant change in level of ionization has occurred on a time scale of a
few orbits. A conservative upper limit for the time of significant XM1/XM3
change is  60 ks, which is roughly the time between the flux level near t=240 ks
and the minimum near t=300 ks. Direct spectral fittings of these two epochs,
assuming \Uox\ variations in direct proportion to the observed flux changes,
are entirely consistent with this idea (see Table 2). They produce acceptable
fits  and confirm that pure continuum slope
variations cannot explain the observed effect (note again the limitation of
spectral fitting that requires integration over many orbits to achieve
adequate S/N).
In particular, the
calculations show that the dominant stage of ionization
is O{\sc vii}, with fractional ionization of about 0.8, and most of the low energy
opacity variations are due to the increasing opacity of O{\sc vi} following the
continuum flux variations. 

The determination of $t_{Ovi}$, the 
time associated with the increase in the O{\sc vi} abundance until it becomes a significant
opacity source (i.e. to exceed a fractional ionization of 0.1),
requires full time-dependent solution of the ionization and recombination equations.
In particular the commonly used 1/e time scale, referred to as ``recombination time'',
is not necessarily a good measure under such conditions.
However,  recombination onto O{\sc vi} provides a good estimate for $t_{Ovi}$,
given  known ionization  equilibrium solution and the fractionl ionization of
O{\sc vi} and O{\sc vii}.
Using these fractions, the
known recombination rates at 
$T \simeq 3.5 \times 10^4$~K 
 (the temperature obtained from the model) and  neglecting
di-electronic and three-body recombination that are not important at these
densities and temperatures, we find
\begin{equation}
 t_{Ovi} \simeq  \frac{ 3 \times 10^{11}  T_5^{0.7} }  {n_e}~{\rm sec}  \,\, ,
\end{equation}
 where $T_5$ is the electron temperature in units of
10$^5$ K. This gives a  lower limit of about 
$2.4 \times 10^6$ \cmiii\  for the recombining gas
density. Combined with the known X-ray luminosity during the 1998 observations,
we derive an upper limit on the distance of the  absorbing gas of 
about $6 \times 10^{17}h_{75}^{-2}$ cm. 
  Given the range in density and
location, and noting the conservative limits on both, the preferred geometry of
the absorbing material is that of a thin shell with $\Delta R/R << 1$

\section{Discussion}

\subsection{Alternative models}

While the  agreement between the $\sim$7 years of 
observations and the model
presented here is satisfactory, we have nevertheless considered various
alternative explanations.

First, we considered the possibility that all observed changes are due
entirely to variations in continuum slope. Spectral fits show that this
requires a slope of  $\Gamma =1.13$ in 2000 and the same \Uox\ as derived for
the 1994 observations (i.e. no change in opacity
between the two epochs). This scenario corresponds to a case where  the
absorbing gas is of very low density, such
that its level of ionization reflects the average luminosity of the source
(which is similar to the 1994 luminosity given the required \Uox). The major
problem  is that, at this ionization parameter, the emission lines produced by
such gas are much stronger and more highly ionized compared with the lines
observed by \chandra. Such lines would easily be detected by the ACIS/LETGS
observations.

A second possibility is that the absorbing gas is indeed responding to the
incident flux yet there are also  large
slope variations associated with the large changes in luminosity.  Slope
variations are claimed to take part in many AGN (Markowitz and Edelson 2001 and
references therein) and are well known in galactic X-ray sources.
It is therefore important to consider their effect on the 1994--2000 spectra of
NGC~3516.

As illustrated by the detailed analysis of the 1998 and 2000 spectra, and as
confirmed by various other tests, we found it practically impossible to
identify  a consistent model with large spectral index variations that fits
the new observations. The main reason is the observational constraints on the
hard X-ray flux. Since  the hard X-ray continuum is directly observed, we can
easily calculate the change in \Uox\ associated with any assumed change in
$\Gamma$. Since the SED and the level of ionization of the gas, during
1994, are known, we can easily calculate the effect such changes will have on
the
line-of-sight gas opacity. As explained, our model assumed no (or very
small) change in slope and hence its predictions represent the
{\it smallest possible changes} in level of ionization. Thus, the
assumption of additional continuum flattening is translated to
a {\it larger change}
in  \Uox\ and hence  larger opacity at
times of lower luminosity. The computations show that such
an 
effect is so large that it will produce a low luminosity spectrum which is
inconsistent with the \chandra\ observations. In fact, as demonstrated
earlier, the largest  change of 0.5--50 keV spectral index that is consistent
with the single absorber model is about $\Delta \Gamma =0.2$.

Another consistency check on the  model  is the
comparison between UV and X-ray observations. NGC~3516 was observed by 
{\it HST}, 
{\it HUT}, and {\it FUSE}\ 
on numerous occasions, 
\centerline{\includegraphics[width=8.9cm]{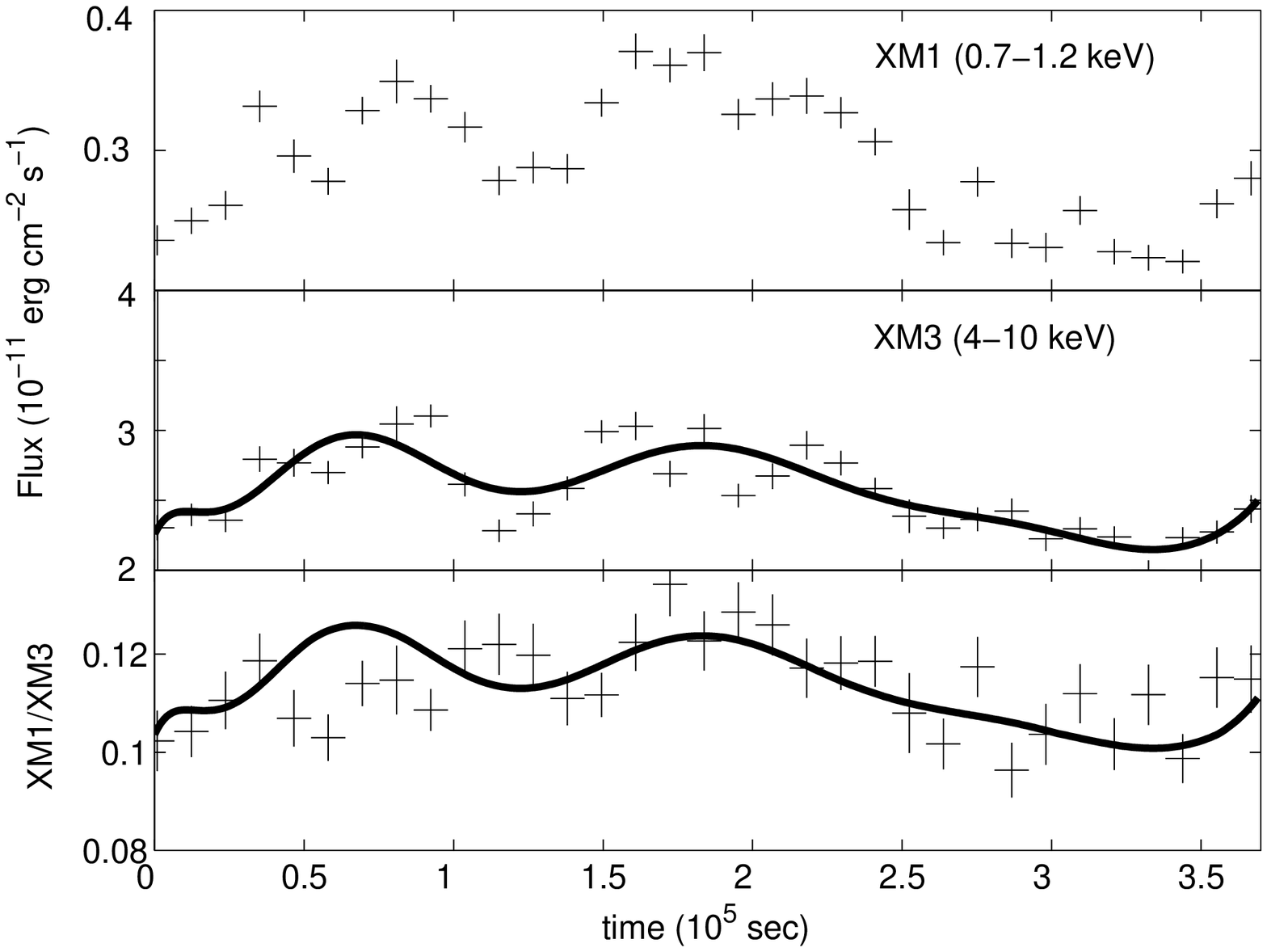}}
\figcaption{ Binned (two-orbits) XM1 and XM3 fluxes,  and XM1/XM3, as
functions of time for the 1998 \asca\ observations.
A smooth model was fitted to the XM3 observations (solid line) and the
model predictions are shown on the XM1/XM3 diagram.
}
\centerline{}
\centerline{}
\noindent
some of these simultaneous with
\asca\ and \chandra\ observations. While a detailed discussion of the UV-X-ray
relationship is deferred to a later paper, we note that the overall ranges in
UV and hard X-ray luminosities is very similar indicating small variations
in the SED.

Finally, combined luminosity, SED and column density variations  
may also be considered. The number of free parameters in such
models is so large that additional constraints, from a large number of
emission and absorption lines, are required to assess their reality.

Several other
papers discussed  the X-ray spectrum of NGC~3516 and suggested alternative
explanations for its spectral variations. Kriss et al. (1996b)  suggested a
range of ionization parameters and column densities. Kolman et al. (1993)
suggest a connection with the BLR and Mathur et al. (1997) discuss the UV-X-ray
absorption connection.
A  recent \sax-based study, reported in Guainazzi et al. (2001), discusses two
epochs during 1996--1997,
separated by 7 months and showing large spectral variations similar to
the ones discussed here. The authors  adopted a completely
different approach by considering both warm and
cold absorbers with different column densities. Their interpretation of the
two different spectra  is very different from ours involving non-equilibrium
situations,   similar to the
 model proposed by Nicastro et al. (1999), and large changes of spectral index.
Their   favorable model 
includes cold absorbing clouds of large columns ($few \times 10^{22}$ \cmii)
 moving
in and out of the line of sight. According to the model, the observed
variations are due to the onset of substantial cold absorption during
times in which the properties of the primary continuum do not change.

There are observational limitations and theoretical problems in the Guainazzi
et al. (2001) interpretation.
The use of absorption edges instead of a self consistent
photoionization model, and the assumption of a large change of slope,
prevented them from testing
the simpler idea that the    apparent changes in spectral index are due to
changes in    the level of ionization of the absorber. 
Visual inspection of their data
(Fig. 1 there)  suggests, infact, that the two \sax\ observations can
 be fitted into the same sequence of ionization vs. luminosity presented
in Fig. 3.

\subsection{Ultraviolet emission and absorption lines}

The model proposed here has
important observational consequences to the UV properties of the source
at minimum light.
Given the model assumptions
we find large column densities of O{\sc vi}, N{\sc v} and C{\sc iv} and
hence strong emission and absorption lines produced by these species during the
\chandra\ observations. The exact values depend on the assumed (unobserved)
SED and will be discussed, together with new HST/STIS data, in 
 a future publication.

Mathur et al. (1997) discussed possible  relationships between 
X-ray and UV absorbers in NGC~3516. Their  general picture is
similar in some ways to the global properties of the source discussed here, 
but the actual values they have derived for the absorbing column and the
ionization parameter are quite different from ours.
Much of the difference is likely due to the limited resolution of
 the  ROSAT-PSPC 1992 data they used for their analysis. They have noticed
a large change of slope in earlier observations of the source but did not
interpret it as due to the change in opacity. 

An interesting aspect of the Mathur \etal\ study is the suggested link between
the EUV and X-ray properties of the absorber. In particular, they studied
the possibility of an outflowing absorbing gas that produce both UV and X-ray
absorption and made a clear prediction of the ``disappearance'' of such gas,
over a period of $\sim 15$ years, due to the constant increase in its level
of ionization. Our \chandra\ observations are in clear contradiction to  this
prediction suggesting that the increased opacity in 2000 may be from the
same gas responsible for the strong absorption a decade earlier.

\subsection{The location, density and mass of the HIG}
 Our results produce the
best limits obtained so far on the density and location of the HIG in type-I
AGN since they are based on the internal consistency of several observations
fitted with a very simple  model over a very large energy range.
 The only important
variable of the model is the ionization parameter which is normalized 
by the observed high energy continuum flux in 1994. As argued above,
small intrinsic spectral index variations are probably seen but their effect is
far smaller than the change due to  absorber's opacity.
Our observations and model fittings
show the absorber to lie well inside the central 1pc and
perhaps much closer in, at the BLR location or perhaps even closer.

There were several previous  attempts to deduce the HIG location in 
AGN. 
Guainazzi \etal\ (1996) used a long-look \asca\ observation of NGC 4051
to claim a significant change in the O{\sc vii} edge, over a time scale of $10^4$ seconds
during which there was no change in the O{\sc viii} edge strength. However,
they did not show a realistic model to explain this change.
Guainazzi \etal\ (1998) and Uttley \etal\ (1999) observed NGC 4051 in a very
low state that allowed them to place a lower limit on the distance of the cold
reflector. Komossa \&  Meerschweinchen (2000) investigated HIG systems in
four radio-loud quasars and NLS1s, using \rosat\ data, and suggested  edge
variations. While the claimed variability is probably
real, the \rosat-PSPC resolution  does not allow accurate modeling of the
absorption features. Moreover, so far there is no positive detection of HIG
with ``typical'' AGN absorption features in any NLS1.  
 Reynolds and Fabian (1995) and Otani
\etal\ (1996) made detailed studies of the absorption features in MCG-6-30-15
and claimed an O{\sc viii} edge change on a time scale of $10^4$ sec. They also
claimed that the O{\sc vii} edge was observed but did not change during the
same time which caused them to propose that  this feature is due to
gas  much further away from the central continuum. While the observed
variations and the associated time scales are possible,  the
interpretation is highly questionable in view of the new \chandra\ and \xmm\ 
observations of this source (Branduardi \etal\ 2000; Lee \etal\ 2001). In
particular, it is not at all clear whether the observed 0.7--0.9 keV
absorption-like features are due to the   O{\sc viii} and O{\sc vii} edges
suggested by Otani \etal. Infact, they may be related to a complex of broad emission 
lines and much narrower and weaker absorption lines (Branduardi \etal\ 2000).
 A more consistent physical model is certainly
needed to understand this source.

Our new findings are of 
great importance for AGN study since they are the first to show clear and
direct changes of the absorber properties in  response to changes in  source
luminosity. The time span is very long (7 years)   and the variability
amplitude  very large (a factor of $\sim 8$). The lack of clear
distance indicators made it impossible, until now, to distinguish between HIG
systems observed in absorption in type-I sources and the ones observed in
emission in type-II sources. Given the deduced location in NGC~3516, and a
plausible geometry of the obscuring medium (torus?) in type-II sources, it is
evident that the system we have observed is very different, in
location and other properties, from the extended X-ray emission line gas in
type-II AGN. Unfortunately, there is no precise mass estimate for this gas. The
combined density and location we derived  give an upper limit to the mass of
about 80 \Msun\ (which is orders of magnitudes larger than the mass of the
ionized gas in the BLR) but much smaller values are entirely consistent with
this estimate.

\subsection{Neutral reflectors and relativistic \Ka\ lines}
Important ingredients in our model are the line and continuum contributions 
 from the low ionization reflector. As explained, this component is necessary to
explain the observed \Ka\ at minimum flux and is consistent with the idea of
a large column-density torus like the ones suggested for 
type-II sources. Interestingly, this component contributes a significant
amount to the observed flux even at maximum (1994 level) luminosity: about 10\%
of the 6.4 keV flux and  EW(narrow \Ka)$\sim 30$ eV under the assumption of
50\% contribution at 6.4 keV during the \chandra\ observations.
This has important consequences  to the intensity and profile of the so-called
``relativistic'' \Ka\ line in NGC~3516 and other AGN.  In particular, the
claimed Fe~K$\alpha$  line in this source (e.g. Nandra et al. 1999) has a typical
broad red shoulder and a sharp blue side. The cold gas contribution, if
similar to that found here, peaks at 5--7 keV and will have a significant
effect on the relativistic line intensity. The detailed analysis of this idea
is beyond the scope of the present work.

\acknowledgments
We are grateful to Niel Brandt and Steve Kahn for useful discussion.
HN acknowledge the hospitality and support of members of the UMBC physics department
where part of this work was performed.
This research is supported by the Israel Science Foundation 
and the Jack Adler Chair of Extragalactic Astronomy at Tel Aviv
University.

\end{document}